\documentclass[prl,aps,twocolumn,floats,nofootinbib,showpacs,superscriptaddress]{revtex4}

\usepackage{amsmath,amssymb,dsfont,graphicx,psfrag}
\usepackage{epsfig}
\usepackage{graphicx}

\begin{document}
\title{Collective Edge Modes near the onset of a graphene quantum spin Hall state}

\author{Ganpathy Murthy}
\affiliation{Department of Physics and Astronomy, University of Kentucky, Lexington KY 40506-0055, USA}

\author{Efrat Shimshoni}
\affiliation{Department of Physics, Bar-Ilan University, Ramat-Gan 52900, Israel}

\author{H.~A.~Fertig}
\affiliation{Department of Physics, Indiana University, Bloomington, IN 47405, USA}

\date{\today}
\begin{abstract}
Graphene subject to a strong, tilted magnetic field exhibits an
insulator-metal transition tunable by tilt-angle, attributed to the
transition from a canted antiferromagnetic (CAF) to a ferromagnetic
(FM) bulk state at filling factor $\nu=0$. We develop a theoretical
description for the spin and valley edge textures in the two phases,
and the implied evolution in the nature of edge modes through the
transition. In particular, we show that the CAF has gapless neutral
modes in the bulk, but supports gapped charged edge modes. At the
transition to the FM state the charged edge modes become gapless and
are smoothly connected to the helical edge modes of the FM
state. Possible experimental consequences are discussed.

\end{abstract}
\maketitle

\textit{Introduction} -- Graphene subject to a perpendicular magnetic
field exhibits a quantum Hall (QH) state at $\nu=0$, made possible by
electron-electron interactions
\cite{Zhang_Kim2006,Alicea2006,Goerbig2006,Gusynin2006,Nomura2006,Herbut2007,
Fuchs2007,Abanin2007,Checkelsky,Du2009,Goerbig2011,Dean2012,Yu2013}. In
particular, the emergence of a $\sigma_{xy}=0$ plateau indicates the
presence of a bulk gap in the half-filled zero Landau level associated
with the formation of a broken-symmetry many-body state. The variety
of different ways to spontaneously break the $SU(4)$ symmetry in spin
and valley space suggests a plethora of possible ground states
\cite{Herbut2007AF,Jung2009,Nandkishore,Kharitonov_bulk,Roy2014,Lado2014,QHFMGexp},
of which the favored one is dictated by the combined effect of
interactions and external fields. Most notably, a phase transition has
been proposed \cite{Kharitonov_bulk,bilayer_QHE_CAF} from a canted
antiferromagnetic (CAF) to a spin-polarized ferromagnetic (FM) state
tuned by increasing the Zeeman energy $E_z$ to appreciable
values. Both phases are in principle accessible in strong, tilted
magnetic fields, and differ in fundamental ways: the CAF is an
insulator, characterized by
gapped charged excitations on the edge \cite{Kharitonov_edge}. By
contrast the FM state
supports gapless, helical,
charged excitations at its edge \cite{Abanin_2006,Fertig2006,SFP,Paramekanti}.  In this work
we address how the edge excitations reflect the differing characters of these states, and
how they continuously evolve into one another as the system passes through the quantum phase transition
between them.

A recent experiment by Young {\it et al.} \cite{Young2013} appears to
manifest the CAF-FM transition in transport measurements, performed in
magnetic fields tilted with respect to the graphene plane.  The key
observation is that for fixed perpendicular component of the total
magnetic field $B_T$, increasing $E_z\propto B_T$ beyond a critical
value $E_z^c$ drives the $\nu=0$ state from an insulator (with
two-terminal conductance $G=0$) to an almost perfect conductor
($G\lesssim 2e^2/h$). This can be attributed to the change in the
corresponding edge states.  In analogy with the quantum spin Hall
(QSH) state in two-dimensional topological insulators
\cite{Kane-Mele,TIreview}, the gapless edge states of the FM state are
immune to backscattering by spin-conserving impurities due to their
helical nature: right and left movers have opposite spin flavors.

In a non-interacting model \cite{Abanin_2006}, the edge modes of the
QSH state are associated with one-dimensional (1D) single-electron
channels, centered at crossing points of dispersing energy levels with
opposite spin index. However, interactions introduce a finite
spin-stiffness and lead to the formation of a coherent domain wall
(DW), and a gap to particle-hole excitations. The low-energy charged
excitations are gapless collective modes associated with fluctuations
of the ground-state spin configuration, in the form of a $2\pi$
rotation in the $(s_x,s_y)$-plane \cite{Fertig2006}. This spin
twist is imposed upon the position-dependent $s_z$ associated with
the DW, thus creating a spin {\it texture}, with an associated charge
that is inherent to QH ferromagnets \cite{QHFM,Fertig1994}.  Gapless
1D modes of the DW (which can be modeled as a helical Luttinger liquid
\cite{SFP}) carry charge and contribute to electric conduction. As
spin waves in the FM bulk are gapped, their interaction with the
gapless edge modes has a minor effect on the 1D dynamics \cite{Mazo2}
and the resulting transport behavior.

In contrast, the CAF phase is characterized by a gap to charged
excitations on the edge \cite{Gusynin2008,Kharitonov_edge}. At the same time, the
broken $U(1)$ symmetry in the bulk (associated with a spin rotation in
the $(s_x,s_y)$-plane) induces a neutral, gapless bulk Goldstone mode.
As described below, a proper description of the lowest energy charged
excitations of this state involves a coupling between {\it
  topological} structures at the edge and in the bulk, associated with
the broken $U(1)$ symmetry. This is particularly crucial in proximity
to the CAF-FM transition, where the bulk stiffness softens and
ultimately controls the energetics of charged excitations.

\begin{figure}[t]
\includegraphics[width=1.0\linewidth]{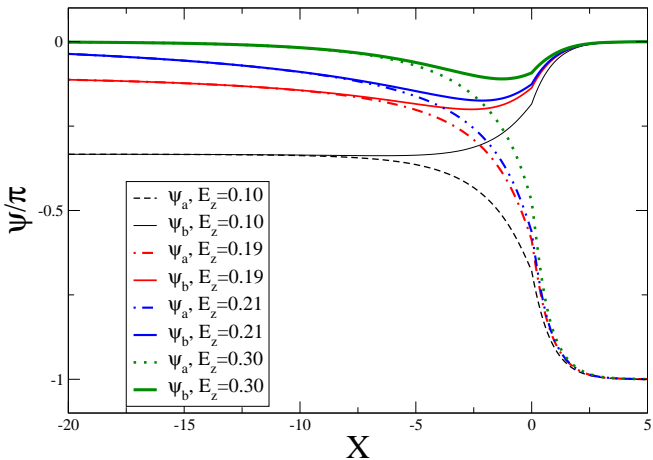}
\caption{(Color online.)  Variation of the canting angles $\psi_a$,
  $\psi_b$ [see Eq. (\ref{aXbX})] with guiding center $X$ (in units of
  $\ell$), in the presence of an edge-potential near $X=0$. The
  critical Zeeman energy for CAF-FM transition is $E_z^c=0.2$.  The
  results are obtained from numerical Hartree-Fock calculations with the maximum edge potential being $U_e=5$ and the width of the edge being $w=3\ell$.  }
\label{fig1}
\end{figure}

In this work, we theoretically describe the evolution of collective
edge excitations as $E_z$ is tuned across the CAF-FM transition. Our
approach significantly generalizes the mean-field ansatz of
Ref. \onlinecite{Kharitonov_bulk} in a way that allows the bulk
and edge of the system
to be treated on an equal footing.
Based upon numerical Hartree-Fock calculations, we derive a
simple description for a spin-valley domain wall configuration at the
edge for arbitrary $E_z$, parameterized by {\it two} canting angles
$\psi_a$, $\psi_b$ (see Fig. \ref{fig1}) which characterize how the
two occupied $n=0$ Landau levels of the $\nu=0$ state evolve as one
approaches an edge.  Low-energy charged excitations can be constructed
by imposing a slowly varying spin rotation on this state.  In the CAF,
these involve binding a vortex (meron) of the bulk state to a spin
twist at the edge, so that the {\it bulk} spin stiffness controls the
excitation energy.  As the CAF-FM transition is approached, the bulk
stiffness vanishes and the vortex unbinds from the edge, yielding a
gapless edge excitation \cite{Fertig2006}.  Our model predicts the
behavior of the activation gap in edge transport as a function of
$E_z$ and offers a qualitative picture of how this transport should evolve
with filling factor. Further experimental
consequences of our model are discussed below.

\textit{Hartree-Fock Analysis and Edge Structure -- } We consider a
monolayer of graphene uniform in the $y$-direction, and subject to an
edge potential $U(x)$ which grows linearly over a length scale $w$ from zero in
the bulk ($x\le0$) to a constant $U_e$ for $x>w$. The system is subject
to a tilted magnetic field of magnitude $B_T$ (dictating the Zeeman
energy $E_z$) and perpendicular component $B_\perp$. The
single-electron states are labeled by a guiding-center coordinate
$X$.  For a given $X$, there are four orthogonal states in the full
$n=0$ Landau level. Their wavefunction can be written in a basis of
4-spinors $|X s\,\tau \rangle$ where $s=\uparrow,\downarrow$ denotes
the real spin index $s_z$, and $\tau=\pm$ is an isospin,
corresponding to symmetric and antisymmetric combinations of valley
states.  The latter are eigenvalues of $\hat{\tau}_x$, and for
convenience our single-particle Hamiltonian implements a simplified
edge potential proportional to this operator
\cite{Kharitonov_edge}. At zero doping, two of the four Landau levels
are filled.

Our model Hamiltonian is projected into the manifold of $n=0$ states labeled by $X$ and has the form
\begin{equation}\label{Htotal}
H=\sum_X c^\dagger(X)
[-E_z\sigma_z\tau_0+U(X)\sigma_0\tau_x] c(X) +H_{int},
%
%
%
\end{equation}
where $c^\dagger(X),c(X)$ are 4-spinor operators,
$\sigma_\alpha$ ($\tau_\alpha$) the spin (isospin) Pauli matrices,
$\sigma_0$ and $\tau_0$ are unit matrices, and $H_{int}$ is the
interaction term. In analogy with Refs.  \onlinecite{Kharitonov_bulk}
and \onlinecite{Kharitonov_edge}, we assume that $H_{int}$ contains
short-range interactions which may break the $SU(4)$ symmetry, but we
also retain an $SU(4)$ symmetric contribution \cite{Yang2006}, so that
$H_{int}$ takes the form
\begin{widetext}
\begin{equation}\label{Hint}
H_{int}=\frac{\pi\ell^2}{L^2}\sum_{\alpha=0,x,y,z}\sum_{X_1,X_2,q}
e^{-q^2\ell^2/2+iq(X_1-X_2)}
g_\alpha :c^\dagger(X_1+{{q\ell^2} \over 2})\tau_\alpha c(X_1-{{q\ell^2} \over 2})
c^\dagger(X_2-{{q\ell^2} \over 2})\tau_\alpha c(X_2+{{q\ell^2} \over 2}):
\end{equation}
\end{widetext}
in which we assume $g_x=g_y \equiv g_{xy}$, $\ell=\sqrt{\hbar c/e
  B_\perp}$ is the magnetic length, $L$ is the system size, and
$:\,:$ denotes normal ordering.  We presume in what follows that the
short range interactions satisfy $g_z>-g_{xy}>0$, which is required to
stabilize a CAF state for small $E_z$ \cite{Kharitonov_bulk}.

Within a set of single Slater determinant (Hartree-Fock) states, with
two states occupied within the four dimensional space for each $X$, we
find numerically that for arbitrary $E_z$ and edge potential $U(X)$
the energy is minimized by a remarkably simple ansatz for the two
filled states, denoted as $|a_X\rangle$ and $|b_X\rangle$:
\begin{eqnarray}\label{aXbX}
|a_X\rangle &=& \cos\left[\psi_a(X)/2\right]|X \uparrow + \rangle -\sin\left[\psi_a(X)/2\right]|X \downarrow - \rangle \\
|b_X\rangle &=& -\cos\left[\psi_b(X)/2\right]|X \uparrow - \rangle +\sin\left[\psi_b(X)/2\right]|X \downarrow + \rangle\; .\nonumber
\end{eqnarray}
$\psi_a(X)$ and $\psi_b(X)$ represent canting angles of the spin,
which vary continuously as a function of $X$, and are generally
different as depicted in Fig. \ref{fig1}. In the bulk ($X\ll 0$ in
Fig. \ref{fig1}) we recover the configuration found in
Ref. \onlinecite{Kharitonov_bulk}: $\psi_a=\psi_b=\psi$, where for
$E_z<E_z^c=2|g_{xy}|$ a CAF is established with $\psi$ obeying
$\cos\psi=E_z/E_z^c$, while for $E_z>E_z^c$ (the FM phase),
$\psi=0$. However, $\psi_{a,b}$ deviate from this uniform solution
near the edge, smoothly approaching an isospin-polarized, spin singlet
state with $\psi_a=-\pi$, $\psi_b=0$ for large $U_e$.  In the
intermediate region, spin and isospin are {\it entangled}. It should
also be noted that for non-trivial canting angles
($\psi_{a,b}\not=n\pi$), a manifold of degenerate solutions exists
with relative phase factors $e^{i\phi}$ between the $s=\uparrow$,
$s=\downarrow$ spin components.  The arbitrary nature of the angle
$\phi$ indicates that there is a spontaneously broken $U(1)$ symmetry
in the mean-field state, and associated gapless Goldstone modes.  In
the case of the CAF state these are gapless spin-wave excitations in
the bulk.  For the FM state, one finds gapless states at the edge,
which moreover can be used to construct gapless {\it charged}
excitations \cite{Fertig2006}.

An interesting aspect of the groundstate results is that the spatial
scale of the edge structure becomes arbitrarily large as $E_z$
approaches $E_z^c$.  This may be understood by evaluating the
expectation value of the Hamiltonian (Eq. \ref{Htotal}) for a Slater
determinant in which states of the form in Eq. \ref{aXbX} are occupied
for every $X$.  Defining $\psi=(\psi_a+\psi_b)/2$ and
$\chi=(\psi_a-\psi_b)/2$, we assume $\psi$ and $\chi$ evolve slowly
with $X$ and perform a gradient expansion, obtaining an energy
functional of the form
\begin{equation}
E_{HF} \approx A(\psi,\chi) + B_\psi(\psi,\chi)(\psi^{\prime})^2+B_\chi(\psi,\chi)(\chi^{\prime})^2.
\end{equation}
For the FM state, $\psi=\chi=0$, so
that we expect these quantities to be small in the
part of the bulk nearest the edge.  Expanding to
quadratic order and dropping an overall constant,
one finds
$A \approx [E_z+2g_{xy}]\psi^2 + [E_z+g_{xy}+g_z]\chi^2$,
$B_{\psi} \approx [g_0+g_z-3g_{xy}]/4$, and
$B_{\chi} \approx [g_0-2g_z-g_{xy}]/4$.
This form of the energy functional implies that
$\psi$ and $\chi$ will decay into the bulk with
length scales
\begin{eqnarray}
\ell_{\psi}=\sqrt{\frac{g_0+g_z-3g_{xy}}{E_z+2g_{xy}}}, \quad
\ell_{\chi}=\sqrt{\frac{g_0-2g_z-g_{xy}}{E_z+g_z+g_{xy}}}. \nonumber
\end{eqnarray}
Note that $\ell_{\psi}\rightarrow\infty$ for
$E_z\rightarrow E_z^c \equiv -2g_{xy}$,
indicating a divergent length scale at the edge as
the bulk transition is approached.  An analogous divergent
length scale is realized on
the CAF side of the transition. Thus,
the CAF phase penetrates into the bulk FM phase from the edge
as $E_z$ is lowered towards the critical value.

\begin{figure}[t]
\includegraphics[width=0.85\linewidth,trim= 80 80 150 80]{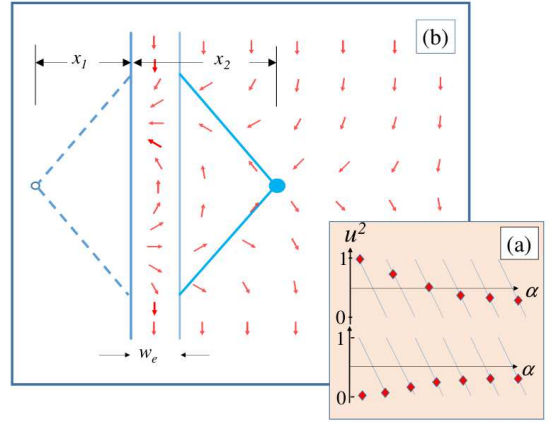}
\caption{(Color online.)  (a) Graphical representation of Hartree-Fock
  wavefunctions in a Landau level supporting a texture.  States of
  different index $\alpha$ (see text) are admixed to implement a
  rotation in $\phi$.  States near $\alpha\rightarrow 0$ must be
  polarized down ($u^2=1$, top) or up ($u^2=0$, bottom).  Position of
  (red) diamond graphically indicates the relative weight of the
  admixed states as $\alpha$ increases.  (b) Model of ``edge soliton''
  consisting of image and real vortices, and in-plane 2$\pi$ spin
  twist at the domain wall.  }
\label{fig2}
\end{figure}

\textit{Charged Excitations: Merons and Edge Solitons} -- Because this
system is a quantum Hall ferromagnet, low-energy charged excitations
may be constructed from slow gradients in the various phase angles
($\psi_a$, $\psi_b$, and $\phi$) in which most of the system is
locally in a groundstate configuration \cite{QHFM,Fertig1994}.
Vortex-like excitations of these systems have non-trivial core
structures and are generically known as merons.  One approach to
evaluating their charge is by explicit construction of wavefunctions
with the appropriate topology.  In the present context these can be
written in the form
\begin{eqnarray}
|\Phi\rangle=
\prod_{\alpha,\alpha^{\prime}}
[u_a(\alpha)c^{\dag}_{\downarrow,-}(\alpha)
+v_a(\alpha)c^{\dag}_{\uparrow,+}(\alpha_+)] \nonumber \\
\times [u_b(\alpha^{\prime})c^{\dag}_{\downarrow,+}(\alpha^{\prime})
+v_b(\alpha^{\prime})c^{\dag}_{\uparrow,-}(\alpha_+^{\prime})]
|0\rangle
\label{skyrmion}
\end{eqnarray}
using a circular gauge, where the
index $\alpha$ represents the angular momentum (integer) quantum number
$m$, and $\alpha_+=m+1$; the coherent combination of $m$ and $m+1$ angular
momenta in the single particle states implements the vorticity of the
in-plane spins \cite{Fertig1994}. (The opposite vorticity
is implemented by coupling the $m$ and $m-1$ states.)  The $u_a$, $v_a$, $u_b$,
$v_b$ coefficients must tend to their groundstate values as
$m \rightarrow \infty$ ($-\sin[\psi_a/2]$, $\cos[\psi_a/2]$,
$\sin[\psi_b/2]$, and $-\cos[\psi_a/2]$, respectively), but for
$m \rightarrow 0$, which fixes the state in the vortex core, for both
the $a$ and $b$ states
one must have either $u \rightarrow 0$ or $v \rightarrow 0$.  The
evolution of the $u$'s and $v$'s with increasing $m$ will be smooth in a low
energy state (and should occur over the very long length scale $\sim\ell_{\psi}$
near the FM-CAF transition), so one may graphically
represent these two possibilities
as depicted in Fig. \ref{fig2}(a).

The transfer of weight
from $m$ to $m+1$ with increasing $m$ leads to a deficit or excess of
charge relative to the groundstate.  For the two examples illustrated
in Fig. \ref{fig2}(a) these charges are $-\sin^2(\psi/2)$ when the
spin is polarized downward for $m=0$, and $\cos^2(\psi/2)$ when
polarized upward. (For states of the opposite vorticity, the signs of
these charges are reversed.)
The two different phase angles $\psi_a$ and
$\psi_b$ can independently be polarized upward or downward in
a meron core.  Thus in the CAF state one finds three possible
charges, $\cos{\psi_a}-\mu$, with $\mu=-1,0,1$, where we have used
the property $\psi_a=\psi_b$ in the groundstate.  Charged excitations
in the bulk of finite energy can be generated by combining meron-antimeron
pairs, for which one readily sees the possible charges are
0, $\pm 1$, and $\pm 2$.

In practice, many transport experiments (e.g.,
Ref. \onlinecite{Young2013}) are dominated by charged edge
excitations, which have lower energy than those of the bulk.  In the
FM state, zero energy charged excitations can be generated by imposing
a slow $2\pi$ rotation of the $U(1)$ variable $\phi({\bf r})$ along
the edge \cite{Fertig2006}.  The state representing this has the same
form as Eq. (\ref{skyrmion}), where the index $\alpha$ becomes the
guiding center coordinate and $\alpha_+=X+2\pi\ell^2/L$.  Since
$\psi_{a,b} \rightarrow 0$ in the bulk for the FM state, the phase
twist only changes the single particle states near the edge, leading to a
zero energy state in the thermodynamic limit.

A key question for this system is: what becomes of this gapless
charged mode as one enters the CAF phase?  The construction described
above leads to a high energy state in this case because the phase
twist alters the state throughout the bulk of the system
($\psi_a=\psi_b \ne 0$ for the CAF.)  This large bulk contribution to
the energy can be eliminated if the edge twist is coupled to a bulk
vortex, along with an image of opposite vorticity outside the system,
as illustrated in Fig. \ref{fig2}(b). To estimate the energy of this
configuration we adopt a simple $U(1)$ energy functional of the form
$E={1 \over 2}\int d^2r \rho_s(x) |\vec{\nabla} \phi({\bf r})|^2$,
with a piecewise constant spin stiffness: $\rho_s(x)=0$ for $x<0$,
$\rho_s(x)=\rho_e$ for $0<x<w_e$ representing the phase stiffness of
the edge structure, of width $w_e$, and $\rho_s(x>w_e)=\rho_b$
representing the bulk stiffness.  Taking $\phi({\bf r})$ to be the
sum of opposed $2\pi$ rotations centered at distances $x_1<0$ and
$x_2>w_e$ from the edge as illustrated in Fig. \ref{fig2}(b), one may
minimize the energy of the configuration with respect to these two
parameters.  In the vicinity of the CAF-FM transition this calculation
yields the result
$$
E_{sol}=\pi \rho_b [\log(\xi_{sol}/\eta) + 0.738],
$$
where $\xi_{sol}=(\rho_e - \rho_b)w_e/\rho_b$ and $\eta$ is a short distance cutoff
indicating the core size of the vortex.

Several comments are in order.  ({\it i}) The size scale $\xi_{sol}$
of this ``edge soliton'' is controlled by the {\it bulk} stiffness
$\rho_b$, which in mean-field theory is proportional to the square of
the in-plane spin magnitude, so that $\rho_b \propto \sin^2 \psi_a$.
Thus the energy of the excitation vanishes as $(E_z^c-E_z) \log
(E_z^c-E_z)$ as the CAF-FM transition is approached.  ({\it ii}) The
charge of the edge soliton is the sum of charges in the domain wall
twist and the bulk meron.  Because of the boundary condition, the
former can only have charge $\pm\cos(\psi_a)$.  Combined with the
possible charges for the bulk meron at $x_2$, the net charge of the
edge soliton is $\pm 1$ or 0.  ({\it iii}) In the limit $E_z
\rightarrow E_z^c$ from below, the charged solitons continuously
evolve into the gapless edge excitations of the FM state, with the
meron portion of the excitation ``evaporating'' as the groundstate
spins become polarized along the total field direction.  Thus the
gapless, charged edge mode of the FM state is continuously connected
to a gapped, charged edge mode of the CAF state.

\textit{Discussion} -- In principle, the charged soliton of the CAF
state should control the activation energy for the quantized Hall
effect observed in transport experiments on the CAF such as those of
Ref. \onlinecite{Young2013}.  The general form of this energy,
$E_{sol} \sim \rho_b \log \rho_b$, indicates that one may learn about
the {\it bulk} phase stiffness of the CAF state via edge-dominated
transport.  This stiffness may be renormalized from the mean-field
behavior used in the analysis above by both quantum and thermal
fluctuations.  An interesting possibility due to the latter of these
is that the bulk should undergo a Kosterlitz-Thouless transition at
some finite temperature $T$, above which $\rho_b$ should jump to zero.
Thus we expect that a sufficiently clean system will display cusp-like
behavior in its diagonal resistance as $T$ passes through the
transition point.  If observed, this would yield direct evidence of the
broken $U(1)$ symmetry inherent to the CAF state.

Further properties of these solitons could be uncovered by studying
how edge transport evolves as a function of doping, which forces them
electrostatically into the groundstate.  At low concentrations these
are presumably pinned by disorder, but at sufficiently high density,
they could undergo a depinning transition, leading to dissipative
transport.  Interestingly, possible signatures of such a
metal-insulator transition as a function of doping are evident in data
presented in Ref. \onlinecite{Young2013}.

The FM phase of this system is a spin Hall insulator
which supports a conductance of $2e^2/h$ in the
$T \rightarrow 0$ limit \cite{Abanin_2006,Fertig2006,SFP} due to
the helical nature of the edge states.  While
measurements \cite{Young2013} do show a transition from a
dissipationless to a dissipative state as $E_z$ is increased,
the (extrapolated) $T=0$ conductance seems to fall short of the ideal
value.
While this could result from the mismatch between
quasiparticle states of a metallic lead and
the highly delocalized gapless charged states of the FM edge,
it is interesting to speculate that intrinsic dissipation may arise from
their interaction with other low energy modes at the edge or
in the bulk.  The relevant mechanism explaining the shortfall in
conductance can be distinguished by four terminal measurements
of the diagonal conductance.

In summary, we have developed a model of the $\nu=0$ graphene edge,
demonstrating that it supports unusual gapped charge solitons in the
CAF state which continuously evolve into gapless excitations as the FM
state is entered.  These excitations can provide information about the
phase stiffness of the CAF state, and should control the low energy
behavior of the system in a variety of situations.

Many open questions remain, such as the full structure of the edge
excitations, both neutral and charged, and the effective theory of the
transition. The authors plan to investigate these issues in future work.

Useful discussions with A. Young, P. Jarillo-Herrero, R. Shankar, and E. Berg
are gratefully acknowledged. The
authors thank the Aspen Center for Physics (NSF Grant No. 1066293) for
its hospitality, and for support by the Simons Foundation (ES). This
work was supported by the US-Israel Binational Science Foundation
(BSF) grant 2012120 (ES, GM, HAF), the Israel Science Foundation (ISF)
grant 599/10 (ES), by NSF-PHY 0970069 and NSF-DMR 1306897 (GM),
and by NSF-DMR 1005035 (HAF).

\end{document}